\documentclass[a4paper]{article}

\usepackage{icrc2013}

\title{ Cosmic-ray leptons, magnetic fields and interstellar synchrotron emission}

\shorttitle{Interstellar synchrotron emission modeling}

\authors{
Elena Orlando$^{1}$ and
Andrew W. Strong$^{2}$,
}

\afiliations{
$^1$ W.W. Hansen Experimental Physics Laboratory, Kavli Institute for Particle Astrophysics and Cosmology, Stanford University, Stanford, CA, 94305, USA \\
$^2$ Max-Planck-Institut fuer extraterrestrische Physik, Postfach 1312, 85741, Garching, Germany\\
}

\email{eorlando@stanford.edu}
\abstract{Interstellar synchrotron emission depends on Galactic magnetic fields and on cosmic-ray leptons.
Observations of radio emission are an important tool for studying cosmic-ray propagation models and interstellar electron spectrum and distribution in the Galaxy. 
We present the latest developments in our modeling of Galactic synchrotron emission with the GALPROP code, including polarization, absorption, and free-free emission. Using surveys over a wide
range of radio frequencies and polarization measurements, we derive constraints on the low-energy interstellar cosmic-ray electron
spectrum, magnetic fields and cosmic-ray propagation models.
 This work is of interest for studies of interstellar gamma-ray emission with {\it Fermi}-LAT, and synchrotron for the {\it Planck} mission.}


\begin{document}
\maketitle

%Begin a section.
%%%%%%%%%%%%%%%%%%%%%%%%%%%%%%%%%%%%%%%%%%%%%%%
\section{Introduction}
Cosmic rays (CRs) propagate in the Galaxy and  generate diffuse radio emission, via synchrotron radiation in Galactic magnetic fields (B-fields). 
Interstellar synchrotron emission extends from a few MHz to tens of GHz and depends  on the CR electron spectrum and distribution in the Galaxy, and on the B-fields. 
Observations of radio emission from our Galaxy are an important tool for studying CR leptons and the B-field. 
Combining these observations and direct measurements of local CRs, we can gain information on CR propagation models and CR electron spectrum in interstellar space.

At higher frequencies, in the microwave bands, free-free and dust emission tend to dominate, and the separation in the data of the different foreground components is problematic. 
 We present here our modeling of both total and polarized synchrotron emission and we show the results. Our study has been performed in the context of CR propagation, considering  other recent observations such as gamma rays and CR direct measurements.
 For a given B-field and propagation model we start with a  CR source distribution and follow the propagation of all the
particles, primary protons and other nuclei, electrons, positrons, taking into account
secondary production, diffusion,  and  energy losses throughout the whole Galaxy.
 We then compare the CR fluxes with
measurements at earth and the synchrotron emission models with radio and microwave data.

%%%%%%%%%%%%%%%%%%%%%%%%%%%%%%%%%%%%%%%%%%%%%%%%%
\section{Interstellar radio emission modelling}

GALPROP  is a software package for modeling the propagation of CR in the Galaxy and their interactions in the interstellar medium (ISM). 
Descriptions of  GALPROP  can be found in \cite{moska2000}, \cite{strong2004}, \cite{strong2007}  and references therein. 
See also Moskalenko et al. (these proceedings) for a recent summary.
This project started in the late 1990's \cite{moskalenko98, strong98}, and since then it has been continuously developed.
It allows simultaneous
predictions of observations of CRs, gamma rays \cite{diffuse2, porter2008} and also synchrotron radiation \cite{orlando2012a,  orlando2012b, strong2011}.
See also
the dedicated website\footnote{http://galprop.stanford.edu/}. 

%%%%%%%%%%%%%%%%%%%%%%%%%%%%%%%%%%%%%%%%%%%%%%%%

\subsection{GALPROP developments}
\begin{figure}
\centering
\includegraphics[width=4cm, angle=0]{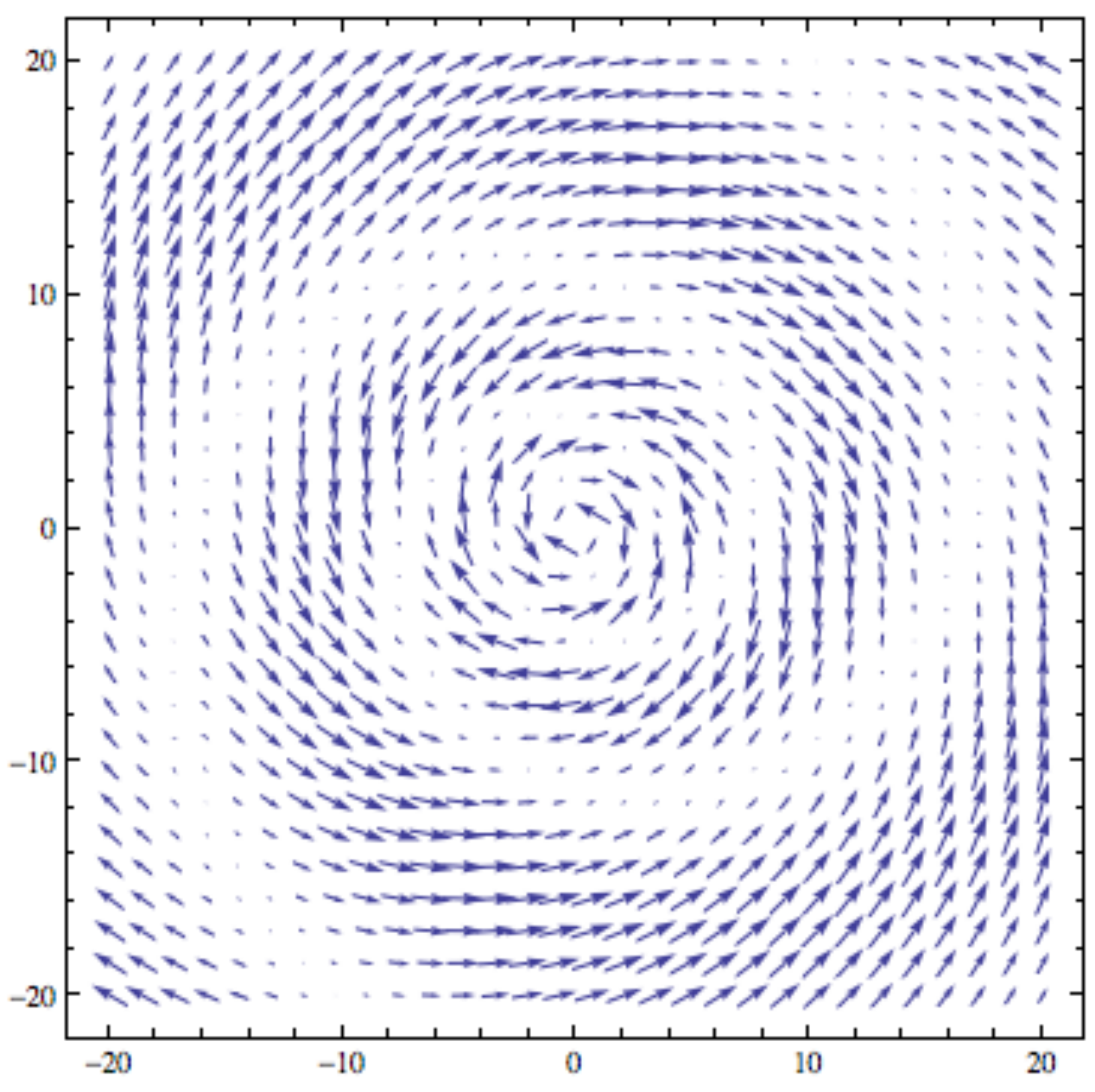}
\includegraphics[width=4cm, angle=0]{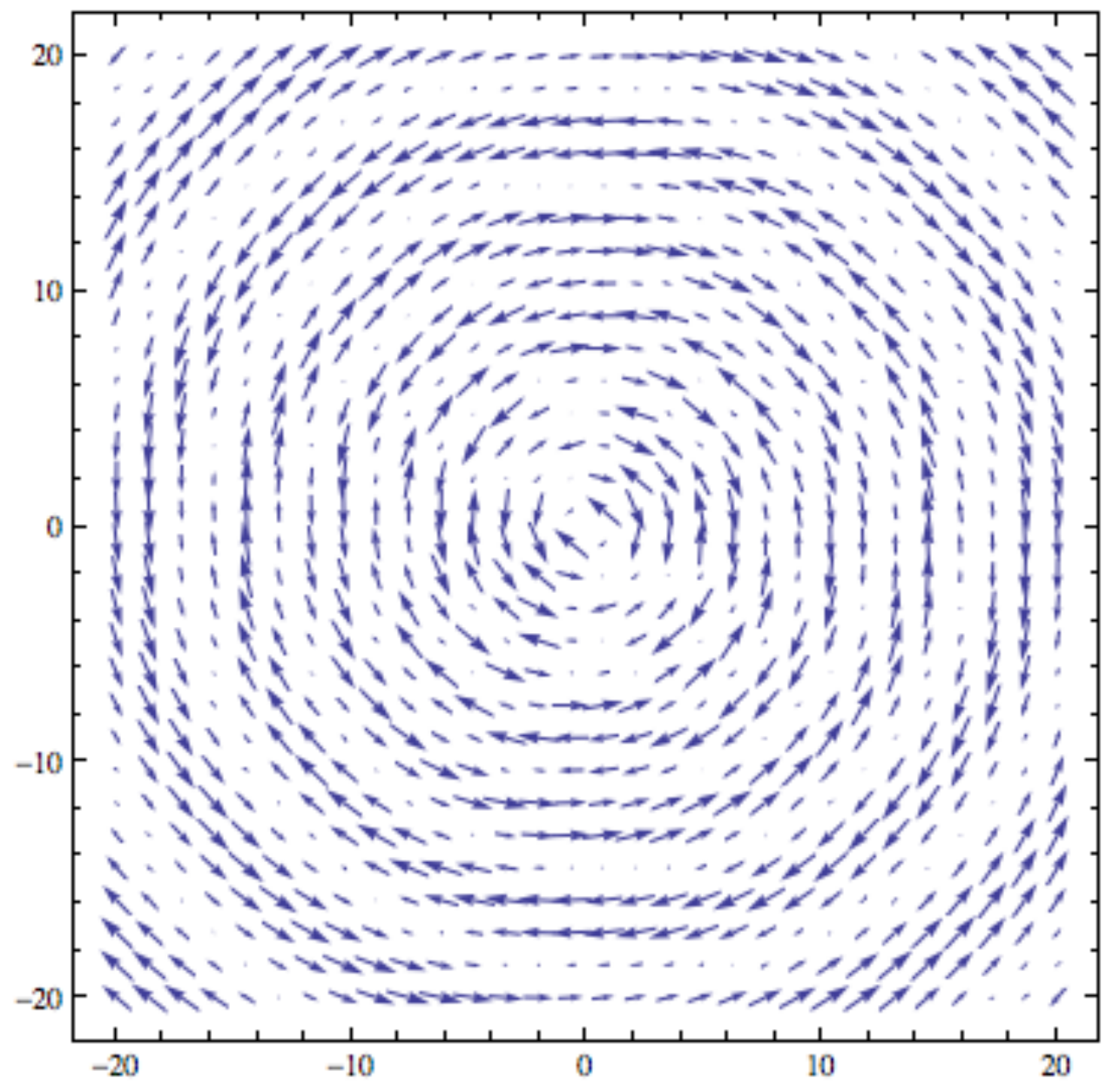}
\caption{ Examples of  B-fields used in our modelling. }
\label{fig1}
\end{figure}

\begin{figure}[h!]
\centering
\includegraphics[width=4cm, angle=0]{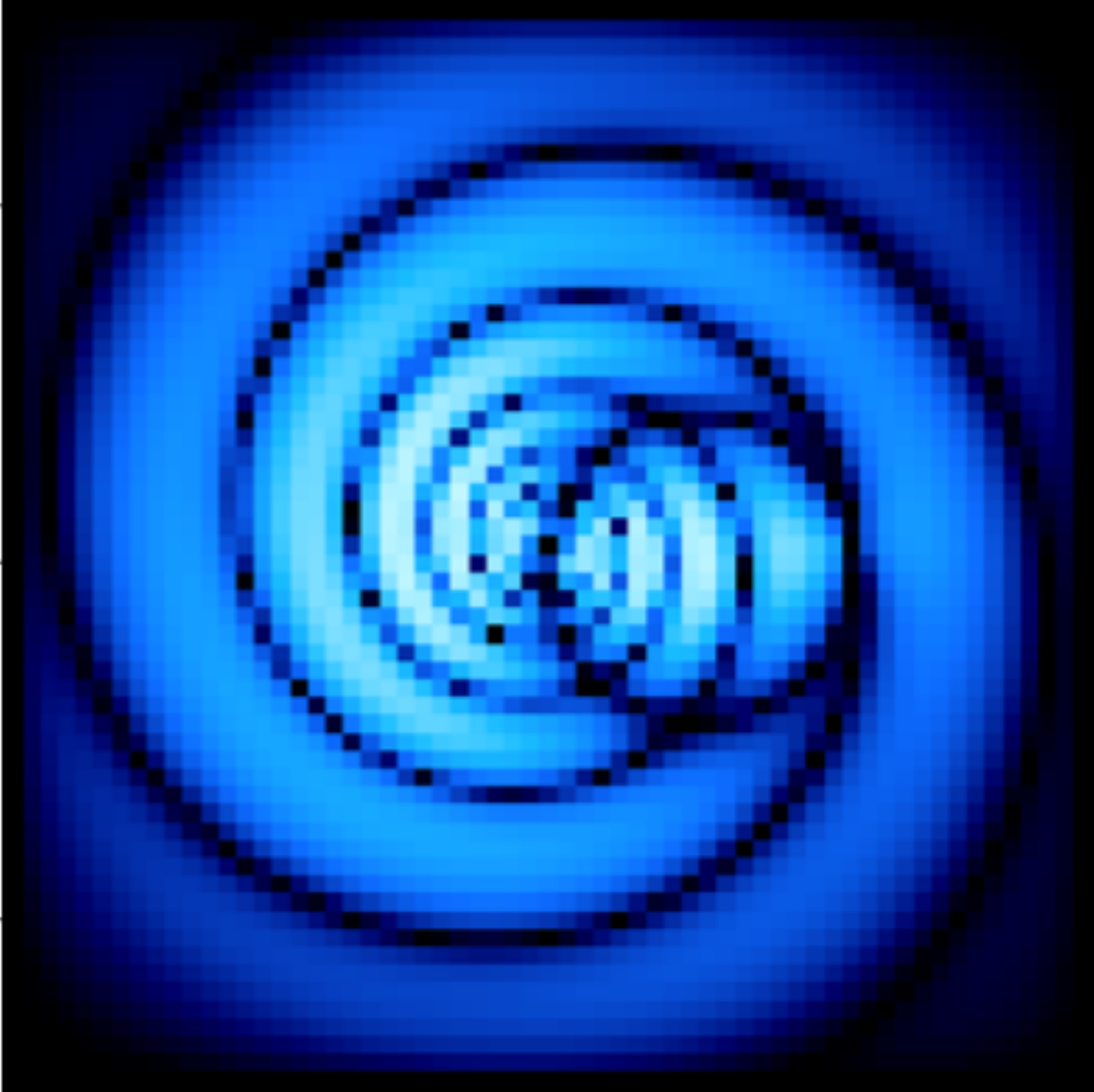}
\includegraphics[width=4cm, angle=0]{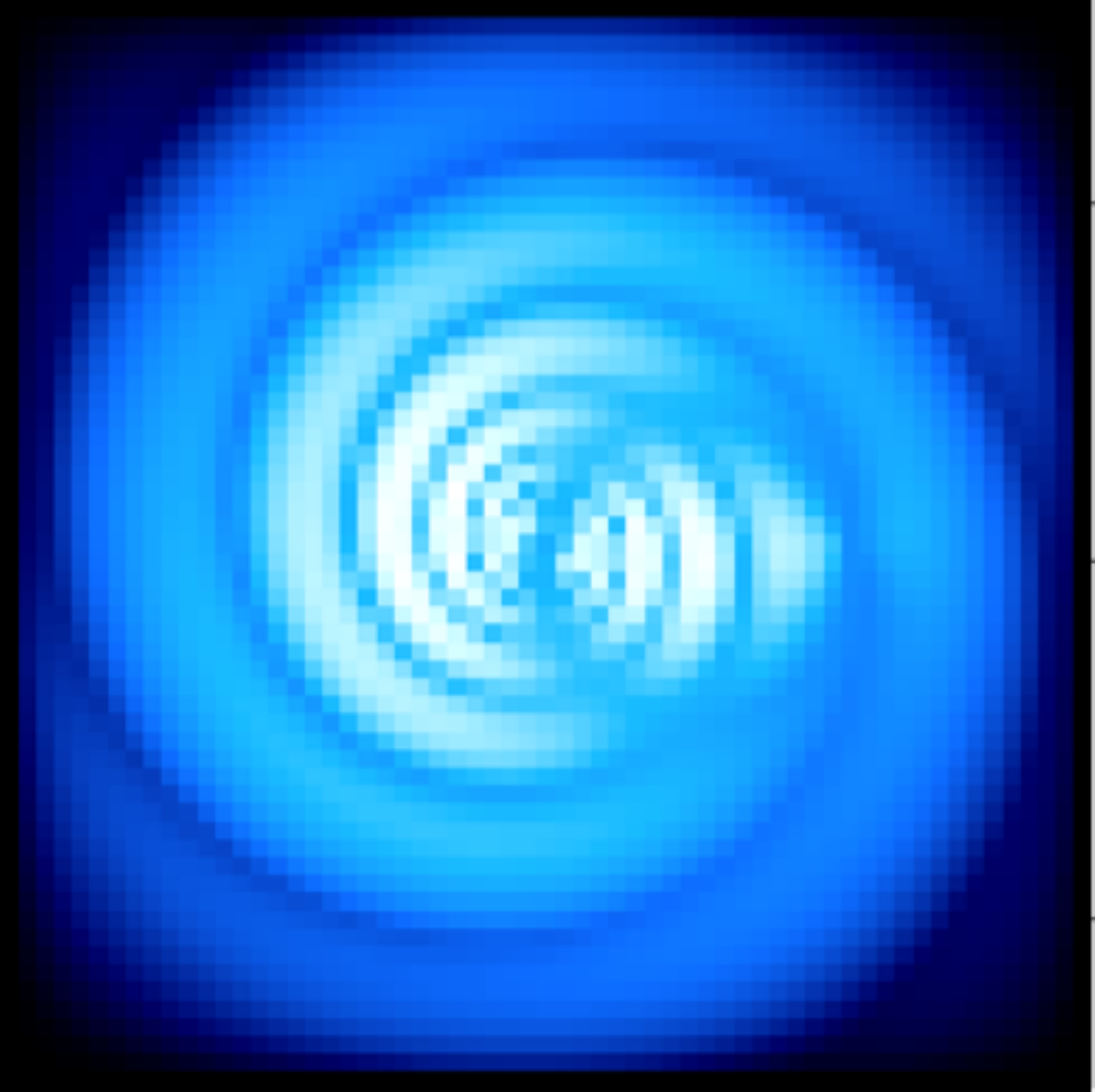}
\caption{ Examples of synchrotron emissivity for regular (left) and total (right) components of our model of magnetic field. The shape of the emissivity reflects the shapes of the spiral arms of the magnetic field. In both figures the Sun is at x=8.5 kpc and y=0. The black circle where the emissivity is zero on the right-center of the figures is where the magnetic field is parallel to the observer's line of sight from the observer point of view in the solar system. }
\label{fig2}
\end{figure}

Recently we have improved the GALPROP calculation  of interstellar synchrotron emission \cite{strong2011}, including 3D models of the B-fields  \cite{orlando2009}, 
and we have also extended it to include   synchrotron polarization and free-free emission  and  absorption \cite{orlando2012a}. More details will be provided in \cite{orlando2012b}.
Implementation of the synchrotron emission model in GALPROP  was described in \cite{strong2011}, where spectra of unpolarized synchrotron emission were presented for a given B-field and for high latitudes.
Different models of the B-field can be implemented; two examples with spiral structure are shown in Figure~\ref{fig1}.
Given a spectrum of electrons or positrons computed at all points on the grid and a  B-field, 
GALPROP integrates
over particle energy to get the synchrotron emissivity.
The emissivity as seen by an observer at the solar position is
computed and output as a function of (R, z, $\nu$)
(for the 2D case) or (x, y, z, $\nu$) (for the 3D case). The spectrum and distribution
of the emissivity depends on the form of the regular
and random components of the magnetic field, and the spectrum
and distribution of CR leptons. All the results presented here were obtained for the 3D case\footnote{See also the Johannesson et al. (these proceedings) for the 3D formulation of CR source distribution.}.
  An example of the emissivities for a given  spiral regular and random B-field component is given in Figure~\ref{fig2}.
To obtain the synchrotron intensity
GALPROP integrates
over the line-of-sight the calculated emissivity on the grid. 
For the polarized synchrotron emission we introduced calculation of the emissivities for Stokes parameters $I, Q, U$ \cite{orlando2012b}.
We integrate the emissivities  over the line-of-sight to produce the corresponding synchrotron skymaps of $I,Q,U$; 
$P$ is computed from the integrated $Q$ and $U$: $P = \sqrt{Q^2+U^2}$. 
The resulting synchrotron skymaps for a user-defined grid
of frequencies are output in Galactic coordinates
either as CAR projection or in HEALPix. 
Skymaps, longitude and latitude profiles, intensity  and spectral index maps can be produced, and the  models can be compared with  radio surveys.

\section{Results}

\subsection{Interstellar electron spectrum}
Below a few GeV the local interstellar electron spectrum cannot be directly measured, because CR electrons with energy lower than a few GeV are affected by solar modulation.  
The synchrotron spectral shape  depends on the CR electron and positron spectra, while the synchrotron intensity depends on the B-field  and electron, positron fluxes.
In particular CR electrons from 500 MeV to 20 GeV produce synchrotron emission from tens of MHz to hundreds of GHz for a B-field of few $\mu$G, and hence this can be used in conjunction with direct measurements to construct the full interstellar electron spectrum from GeV to TeV. 

\begin{figure}[!t]
\centering
\includegraphics[width=15pc, angle=0]{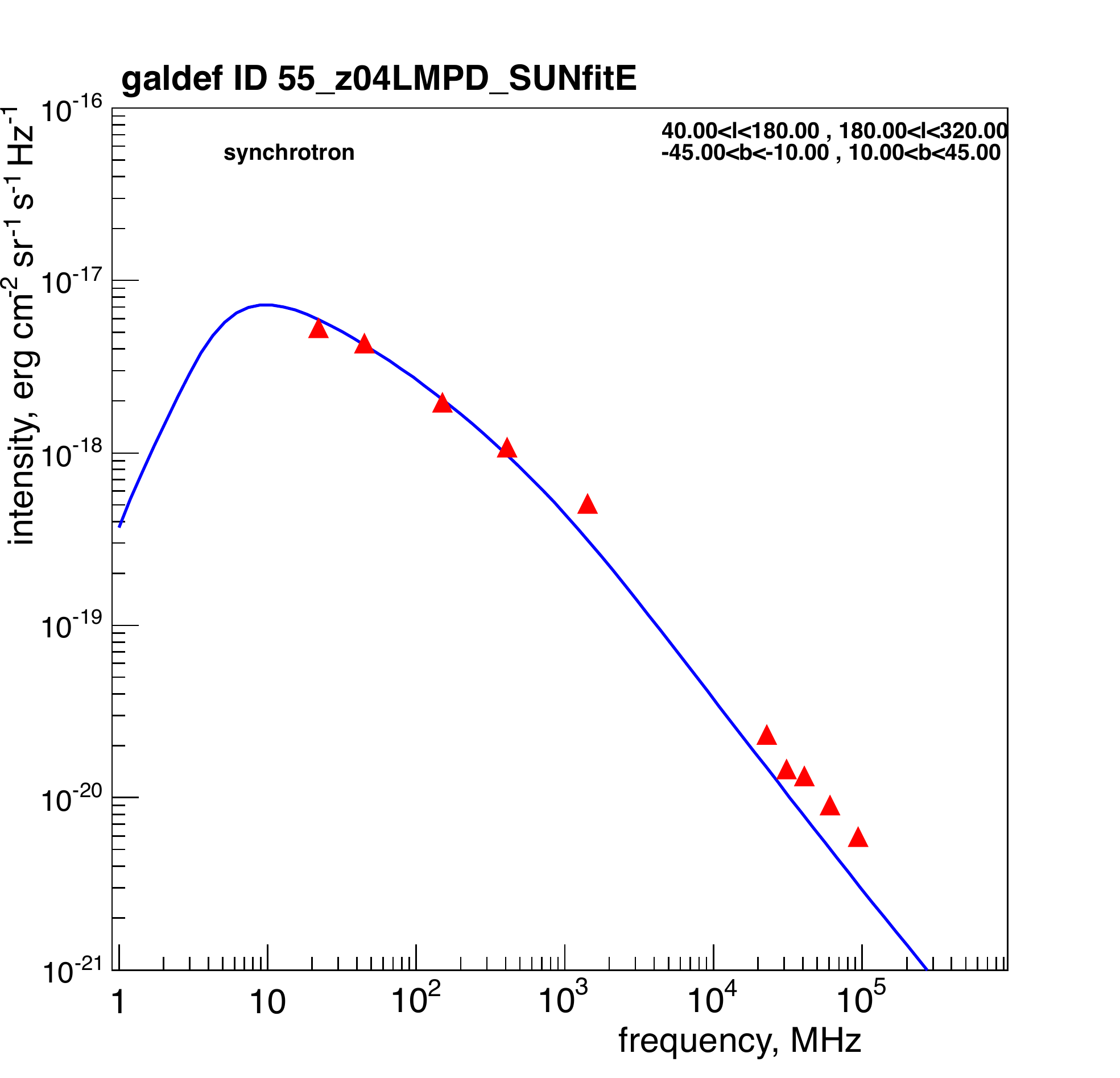}
\includegraphics[width=15pc, angle=0]{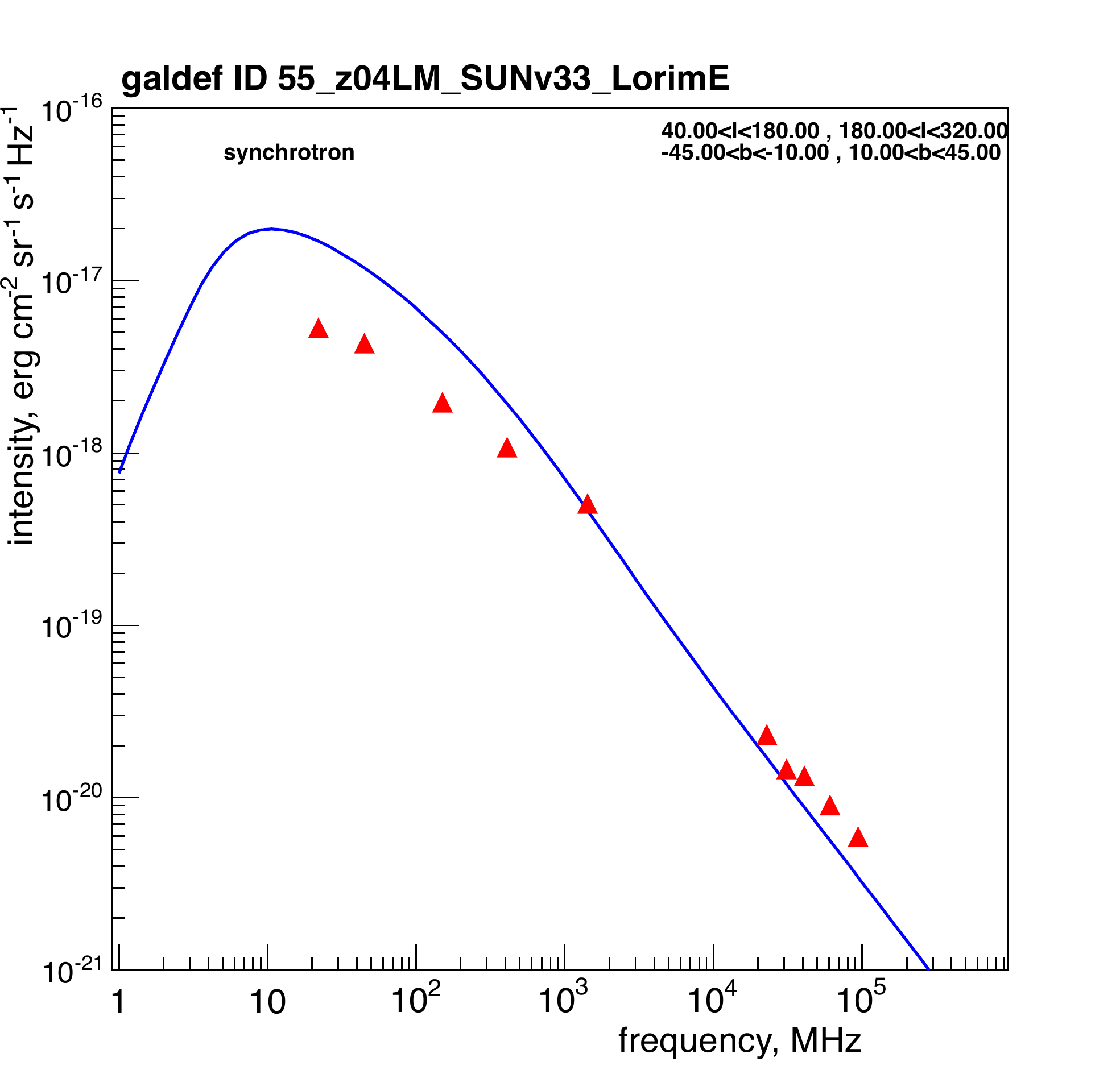}
\caption{Synchrotron spectra (blue lines) with primary low-energy electron injection index 1.6 for pure diffusion model (upper), for diffusive acceleration model as in \cite{diffuse2} (lower) at high latitudes. Data (red triangles) are from radio surveys  and WMAP. }
\label{fig3}
\end{figure}

Using a collection of radio surveys and WMAP 3-year data, out of the Galactic plane, we \cite{strong2011} found that the local interstellar electron spectrum turns over below a few GeV, with  spectral index  3 above the break and around 2 below the break, and harder than 2 below the break. The need of a break in the local interstellar spectrum has been recently observed and confirmed by Voyager (see these proceedings). 
We tested propagation models, generated with GALPROP, in order to constrain the injected CR electron spectrum, before propagation. We found  an injection electron spectral index harder than1.6 below 4 GeV (see Fig~\ref{fig3}, upper plot).

Another result was the recognition of the importance of including secondary positrons and electrons in synchrotron models.
While plain diffusion models fitted the data well, standard reacceleration models used to explain CR secondary-to-primary ratios were not consistent
 with the observed synchrotron spectrum, since the total intensity from primary and secondary leptons exceeded the measured synchrotron emission at low frequencies, as shown in Fig~\ref{fig3} (lower plot).

\subsection{Modeling of synchrotron for various B-fields } 
\begin{figure}
\centering
\includegraphics[width=0.4\textwidth, angle=0] {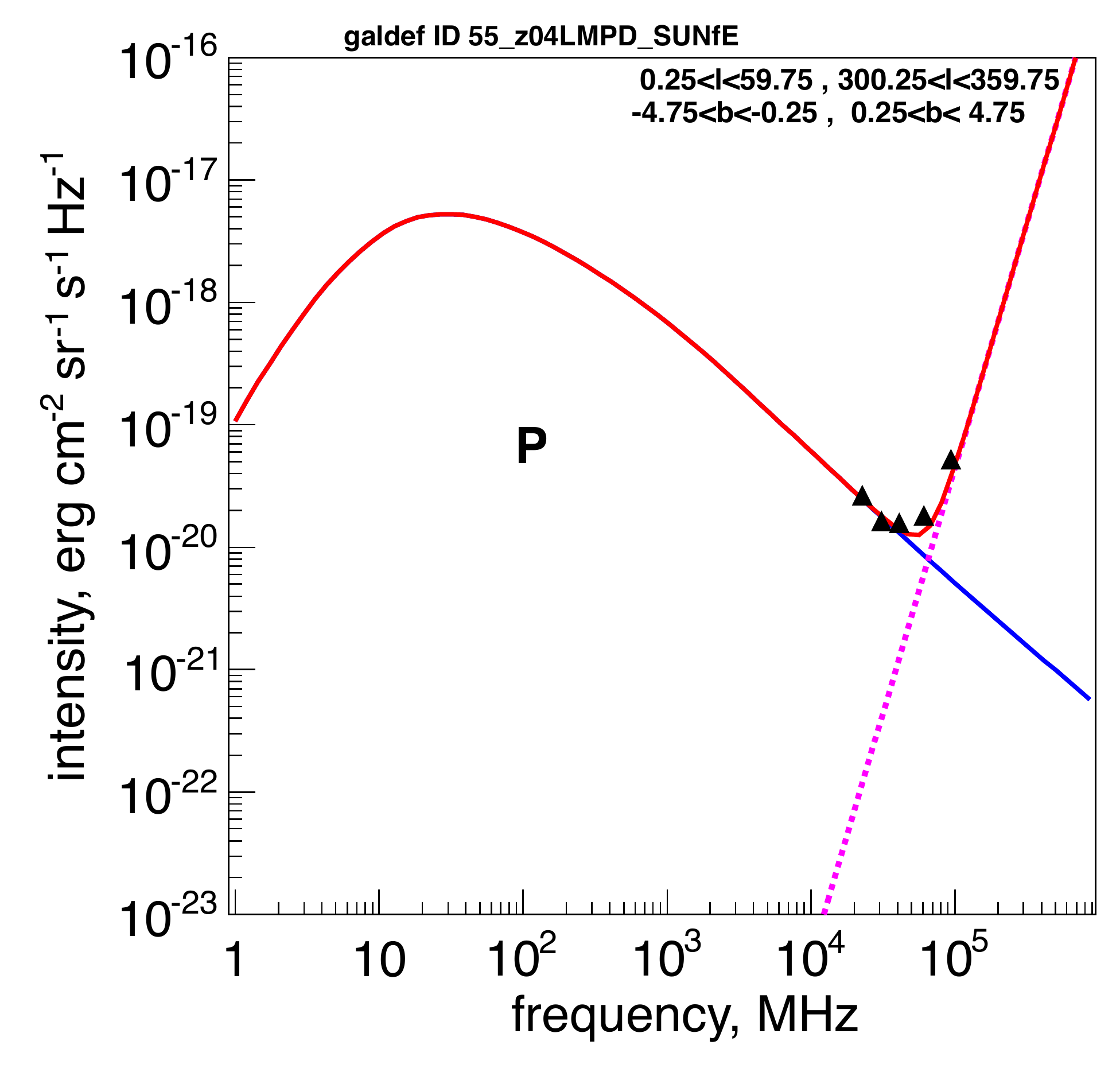}
\includegraphics[width=0.4\textwidth, angle=0] {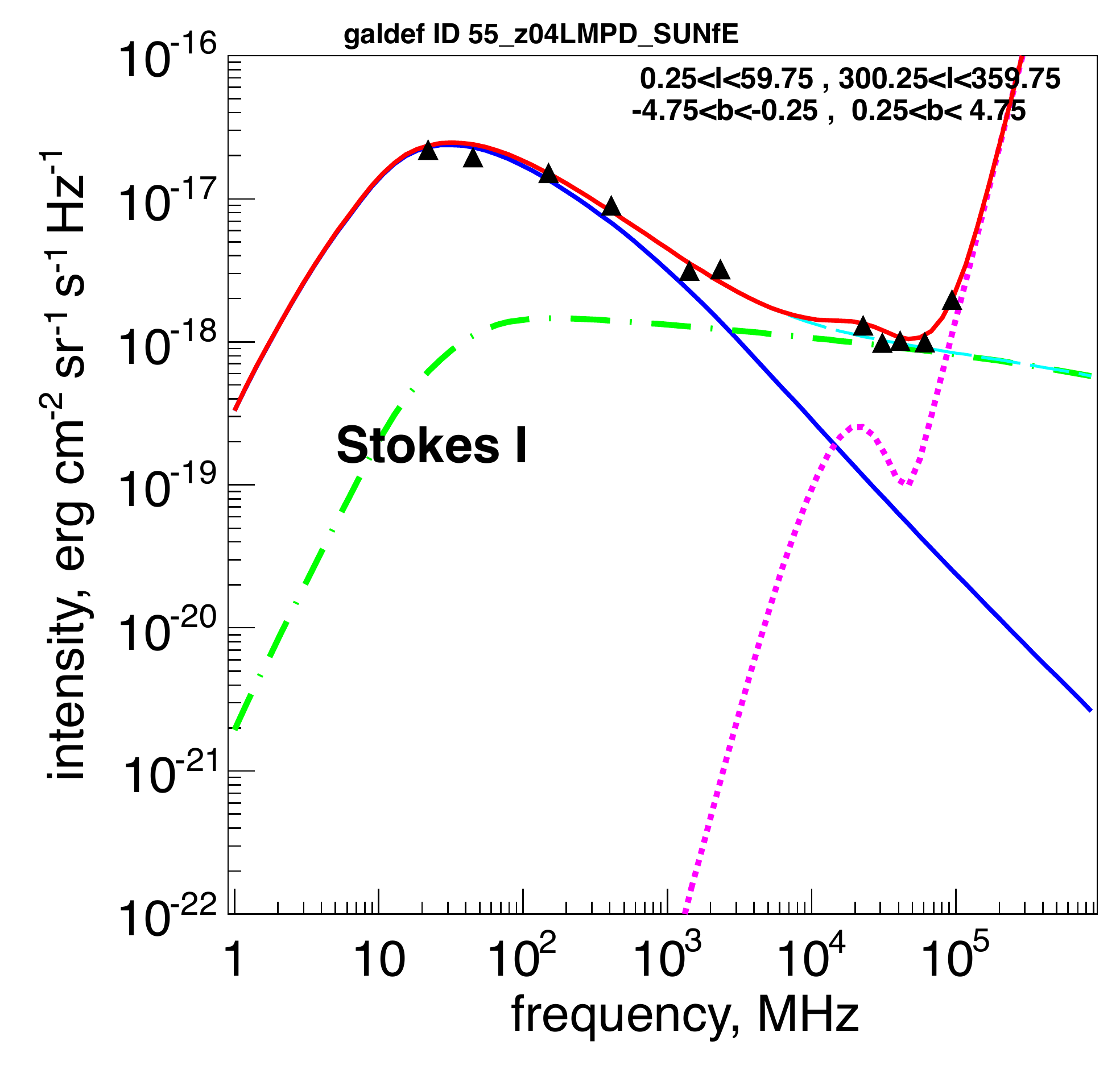}
\caption{  Spectra P and I  for the inner Galaxy and z=4 kpc. B-field intensities are including an anaisotropic B-field to agree with data. Regular B-field model, ASS+RING, based on \cite{sun2008}, for CR source distribution from \cite{strong2010}. The plots show the different model components: synchrotron (blue line), dust and spinning dust (pink dotted line), free-free (green dashed-dotted line), free-free+synchrotron (cyan dashed line) and total (red line). Data (black triangles) are from radio surveys  and WMAP.}
\label{fig4}
\end{figure}

\begin{figure*}
\centering
\includegraphics[width=0.38\textwidth, angle=0] {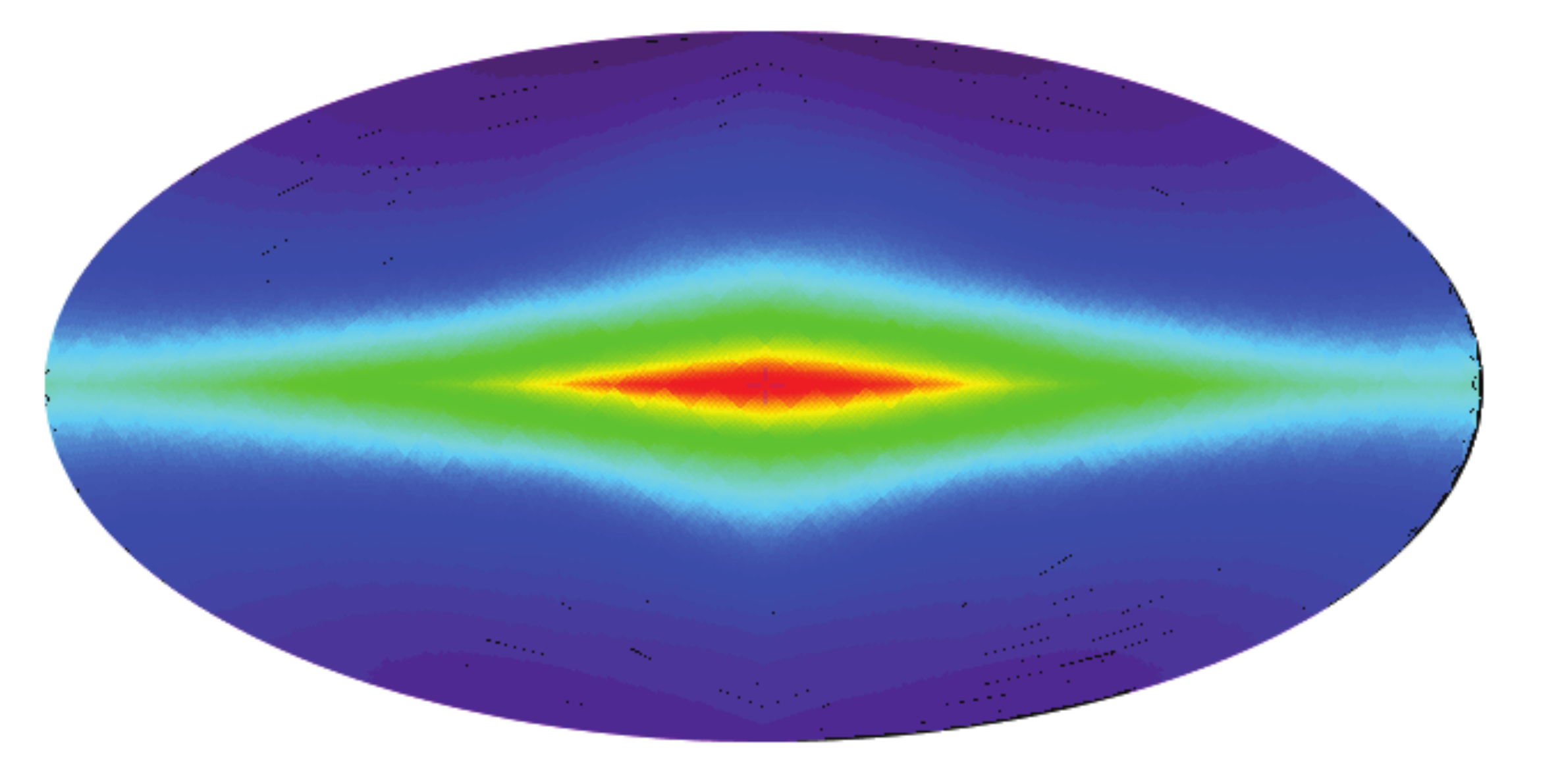}
\includegraphics[width=0.38\textwidth, angle=0] {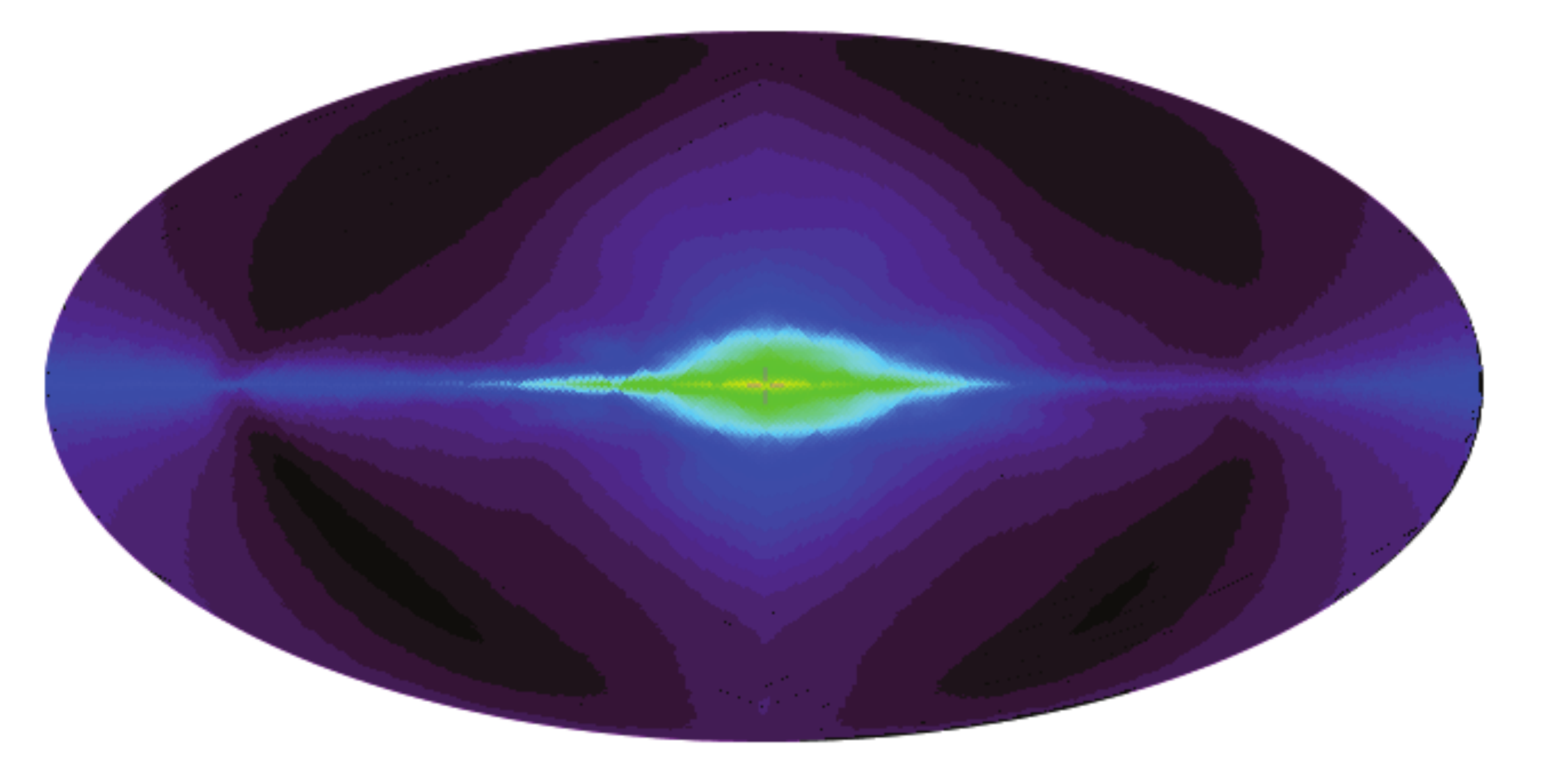}
\includegraphics[width=0.43\textwidth, angle=0] {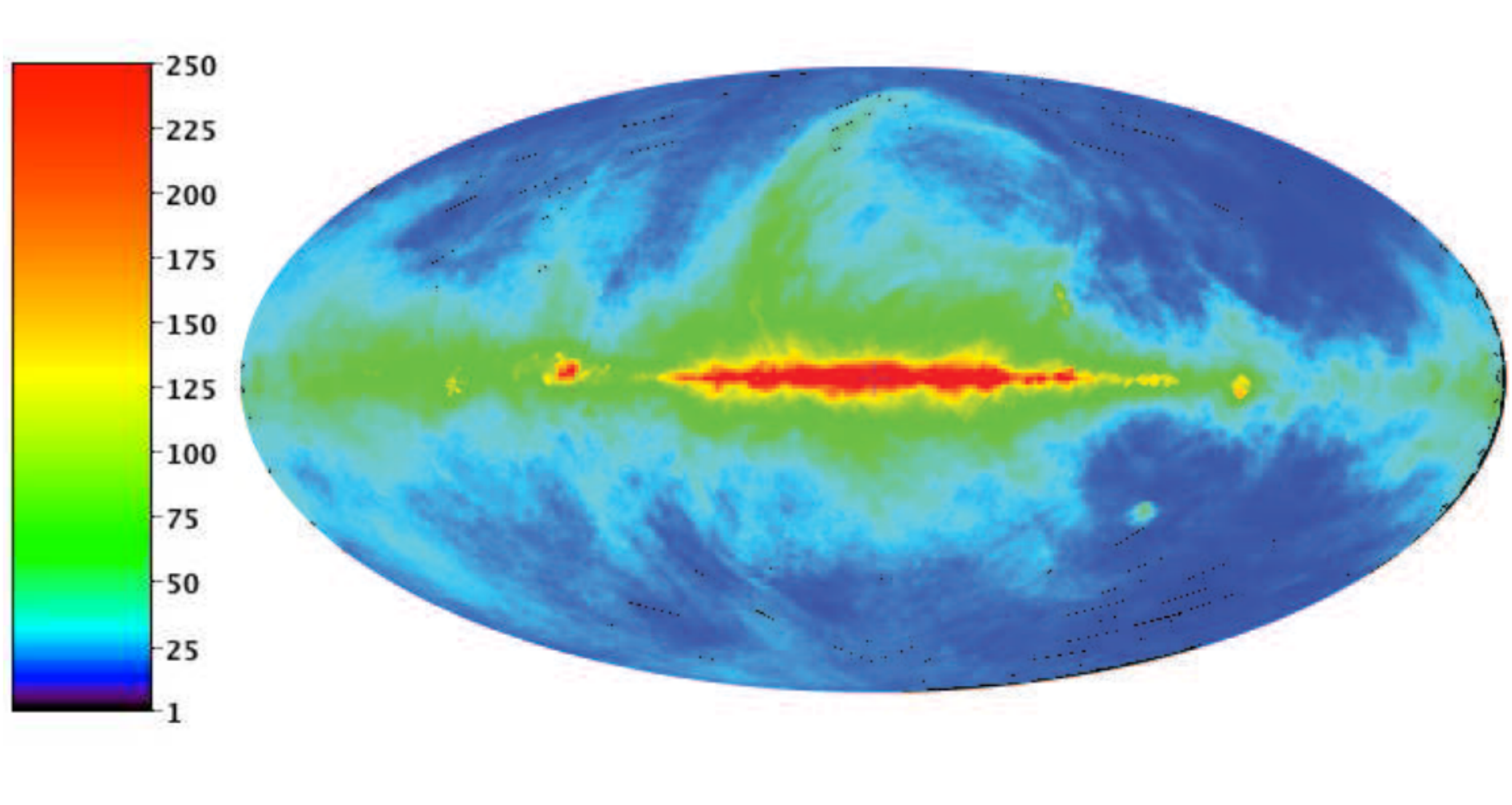}
\includegraphics[width=0.43\textwidth, angle=0] {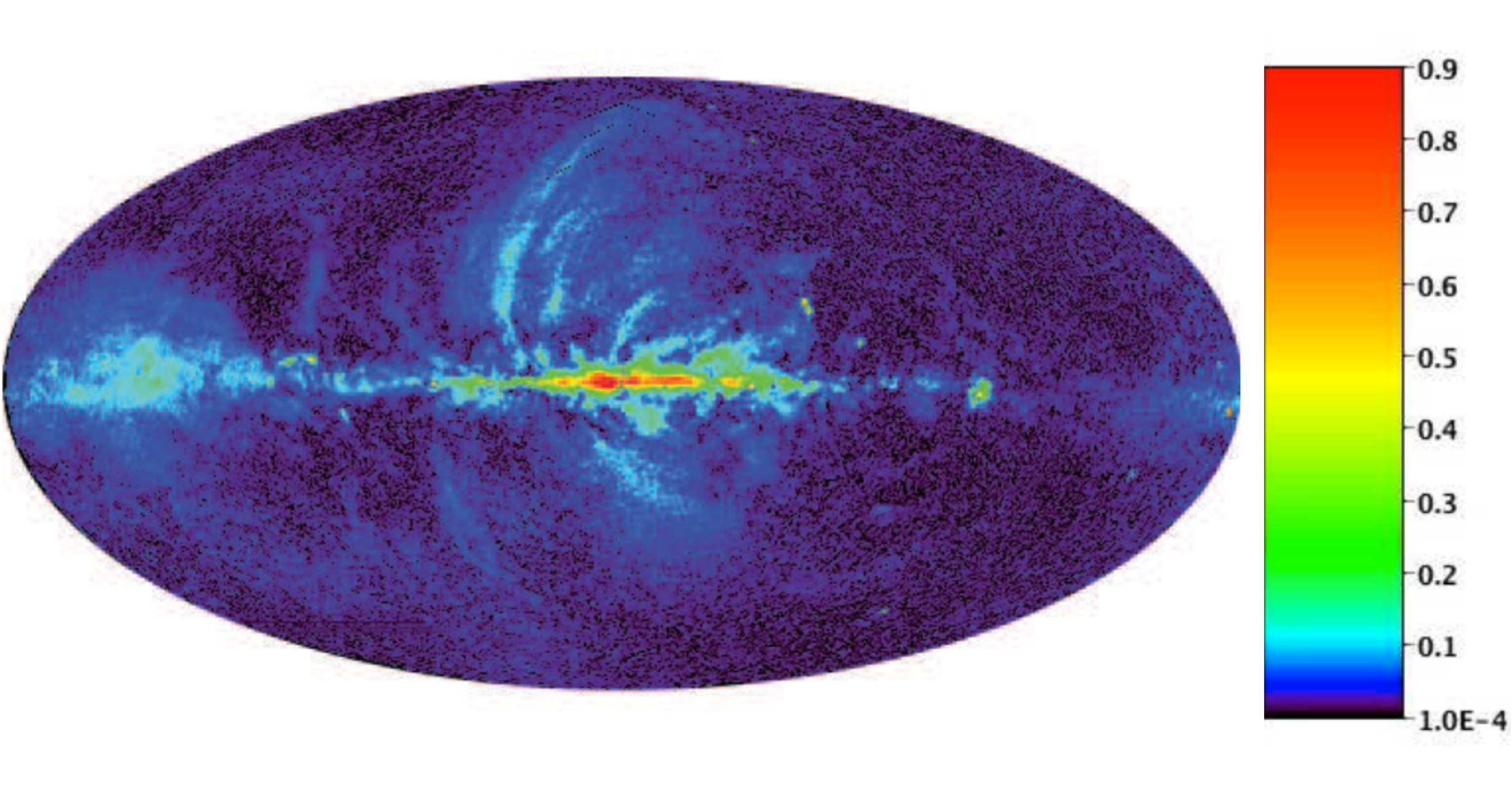}
\caption{Brightness temperature skymaps at 408~MHz  (left) and polarized  23~GHz (right). Top: models, bottom: data from \cite{haslam81} and WMAP. All skymaps (for each frequency) have the same scale, with  Galactic longitude l = 0 at the center. Units are K. From \cite{orlando2012a}.}
\label{fig6}
\end{figure*}

\begin{figure*}
\centering
\includegraphics[width=0.3\textwidth, angle=0] {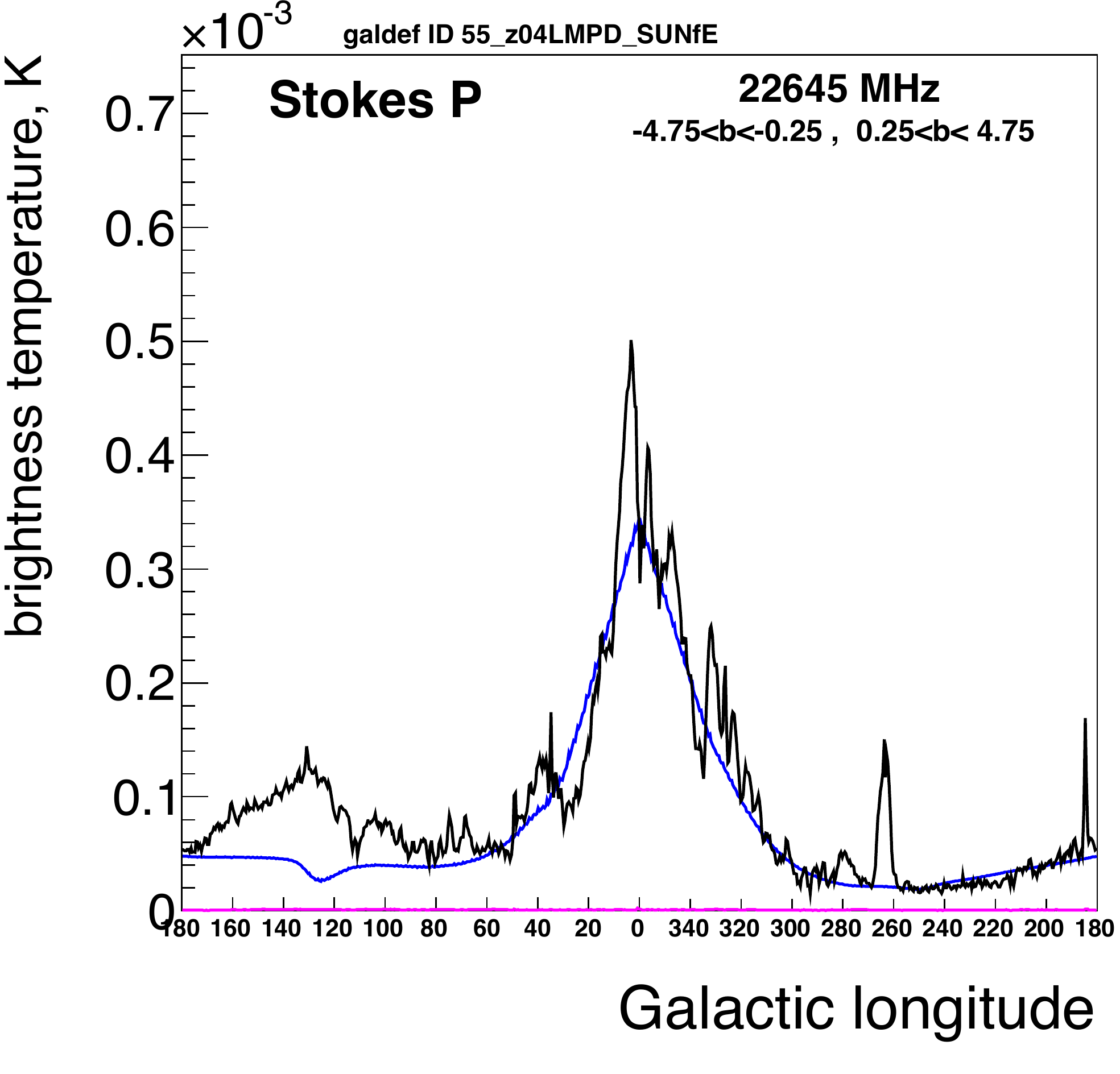}
\includegraphics[width=0.3\textwidth, angle=0] {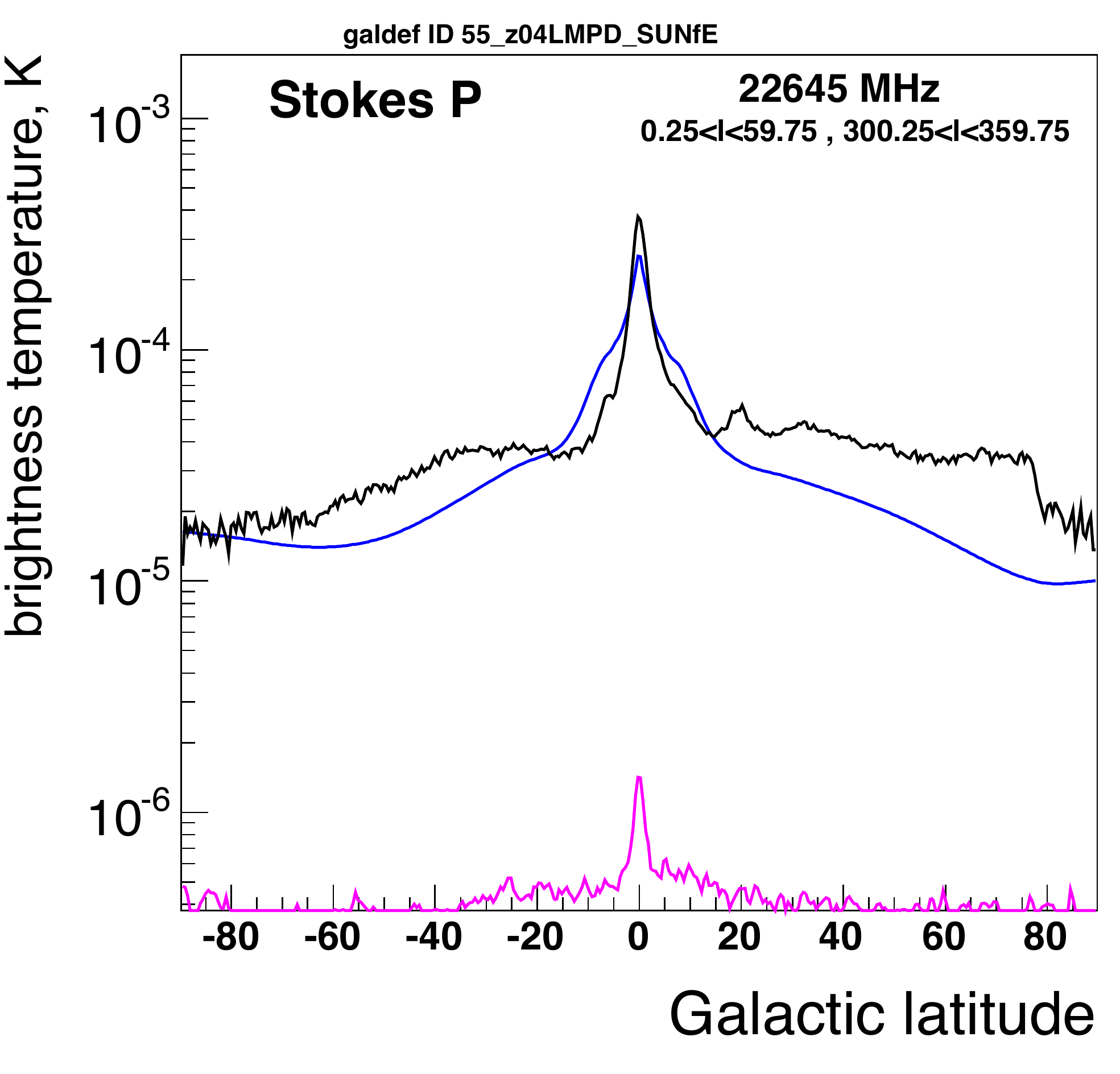}
\includegraphics[width=0.3\textwidth, angle=0] {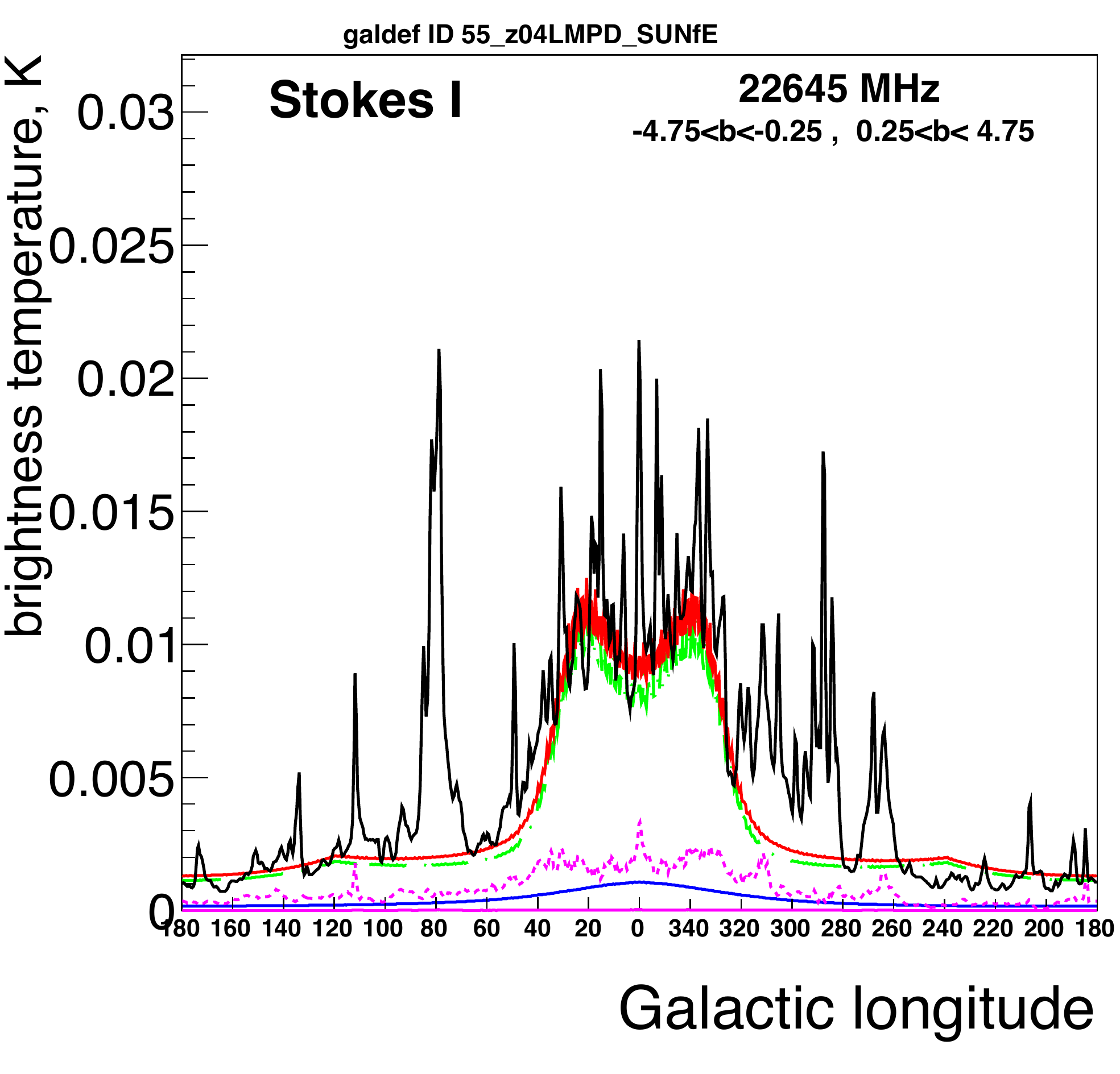}
\includegraphics[width=0.3\textwidth, angle=0] {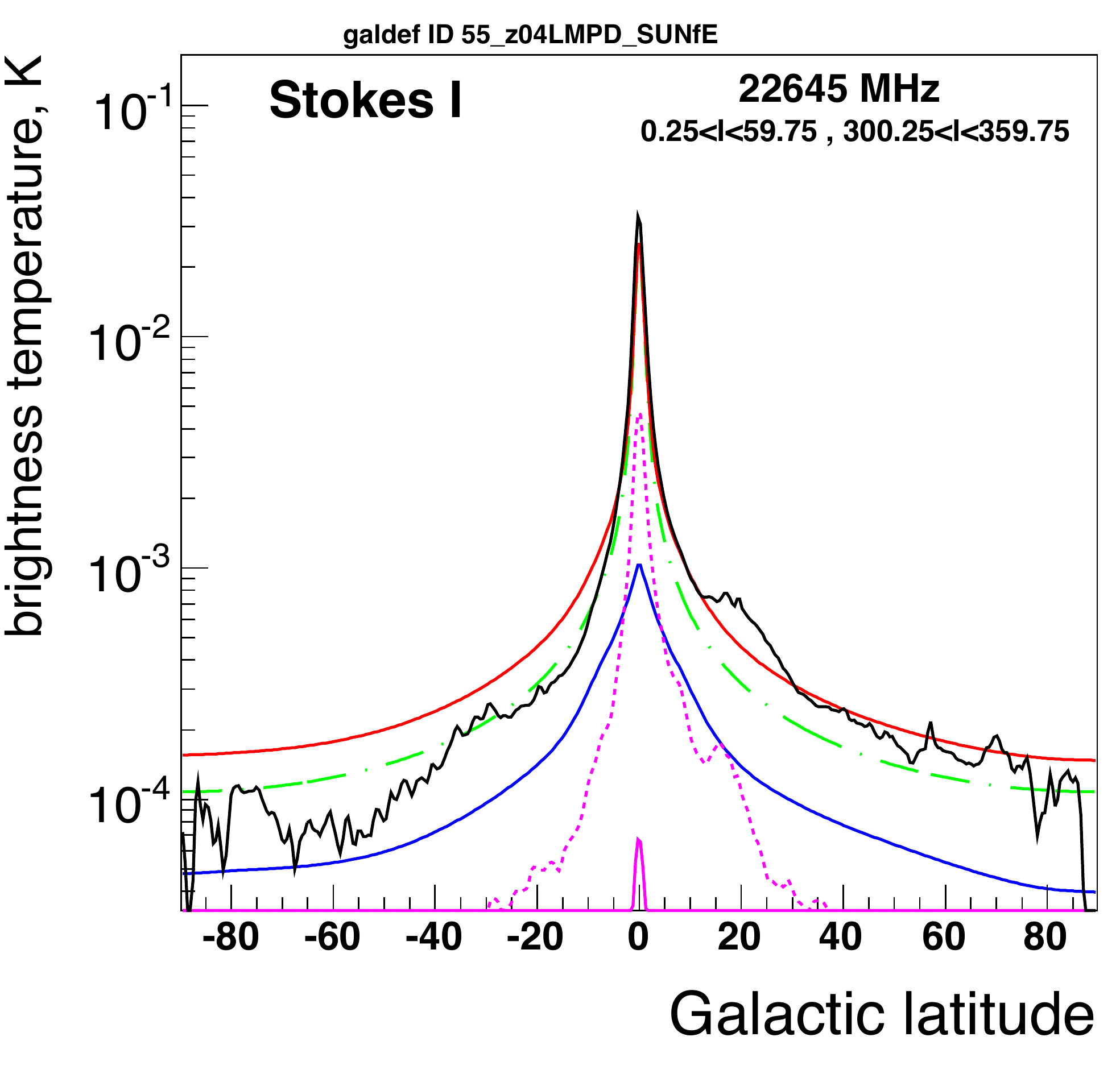}
\includegraphics[width=0.3\textwidth, angle=0] {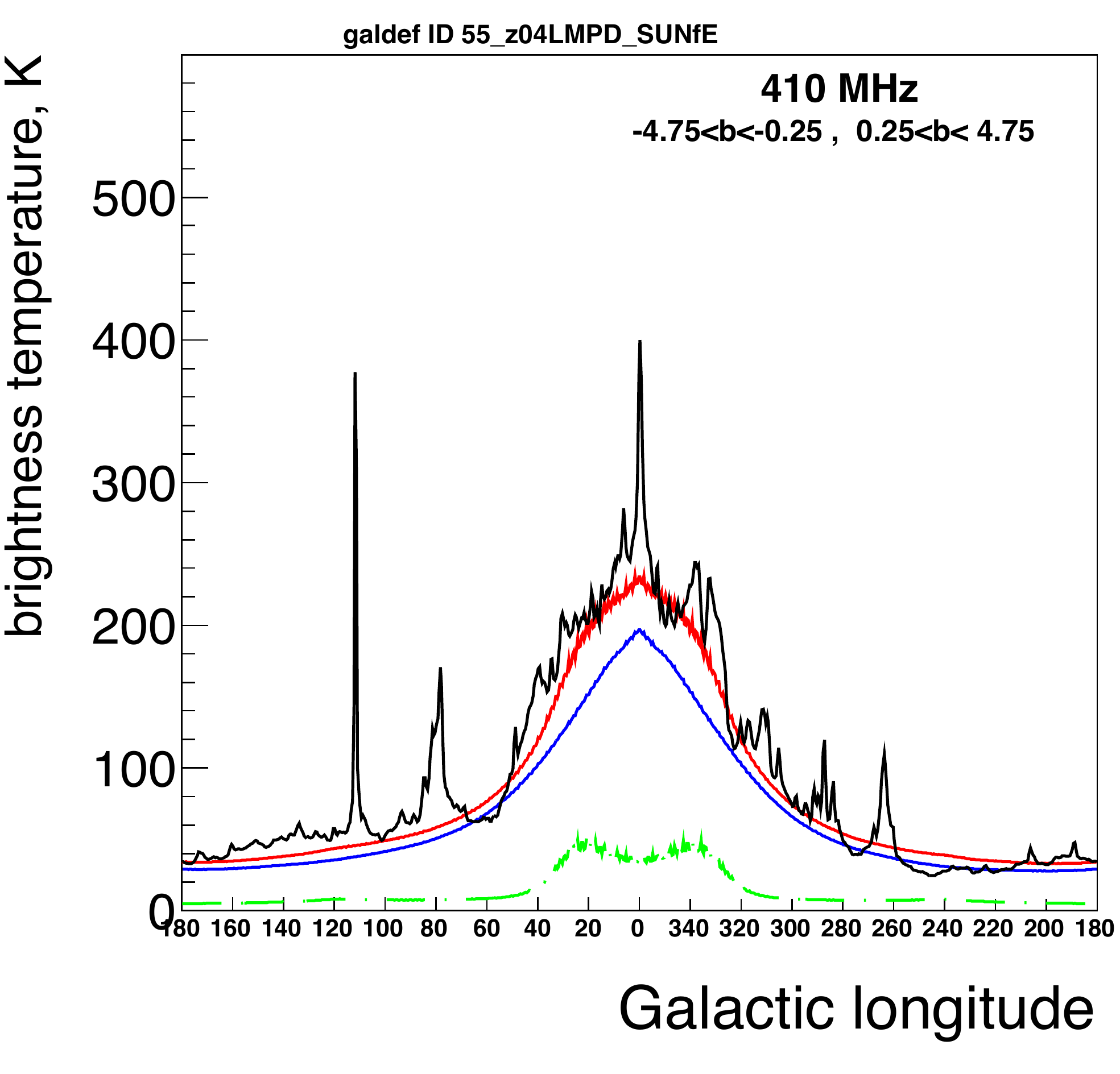}
\includegraphics[width=0.3\textwidth, angle=0] {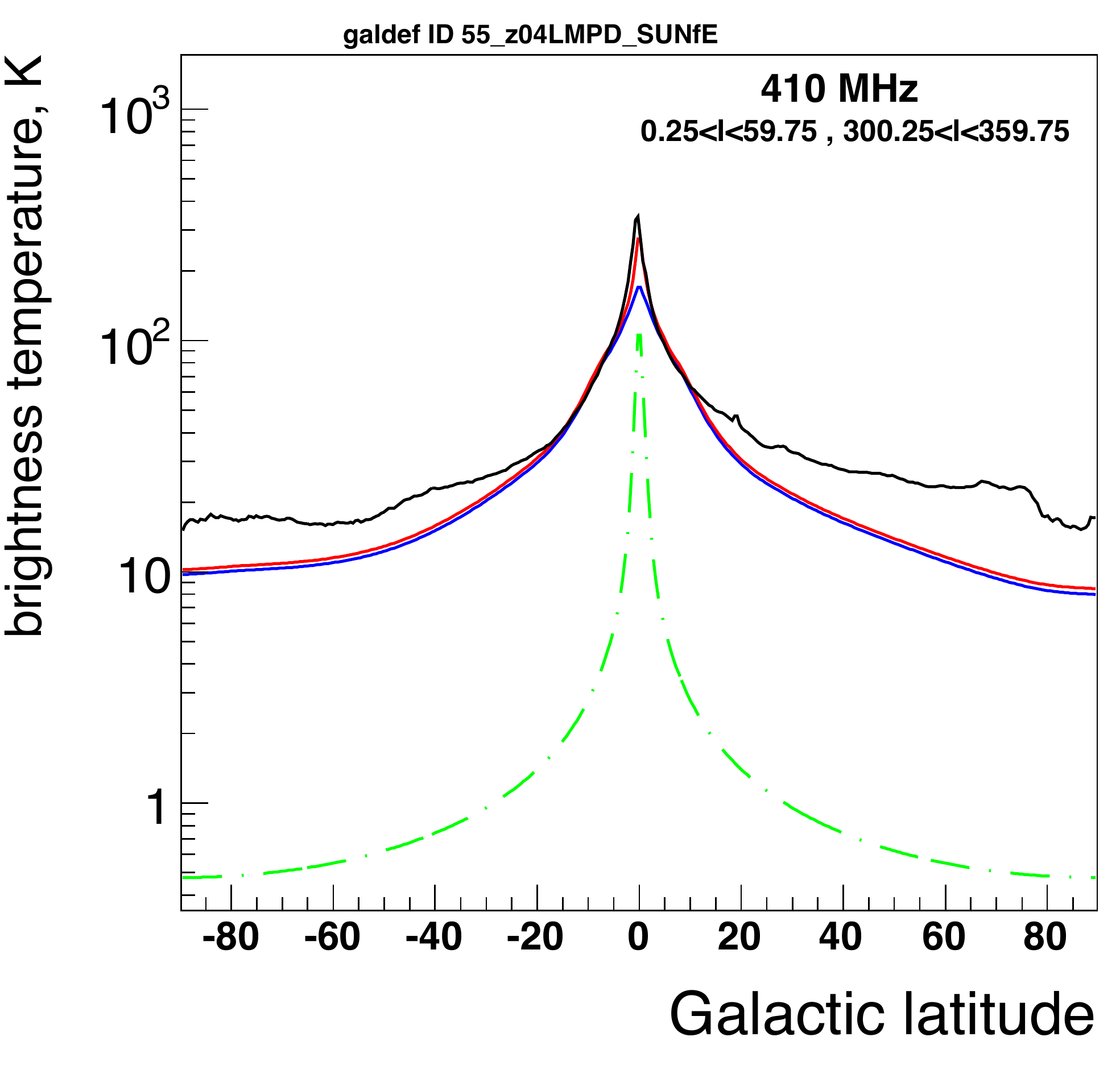}
\caption{ Intensity longitude and latitude profiles for B-field model describe in Figure \ref{fig4}. The plots show the different model components: synchrotron (blue line), dust and spinning dust (pink dotted line), free-free (green dashed-dotted line), free-free+synchrotron (cyan dashed line) and total (red line). Data are in black.  Left to right and top to bottom are longitude and latitude profiles for  P and I at 23 GHz, and 408 MHz. Longitude profiles are averaged over a region of $|b| \leq 5^\circ$, while latitude profiles over a region of $-60^\circ \leq l \leq 60^\circ$ around the Galactic center. From \cite{orlando2012b}.}
\label{fig5}
\end{figure*}

Polarized synchrotron emission allows the regular component of the B-field to be studied, complementing the total synchrotron which probes the sum of random and ordered fields (regular plus anisotropic random).
Models of the regular B-field from the literature were investigated, using CR propagation models based on CR and gamma-ray data. An example of our modelling of the radio emission is shown on Figure \ref{fig4}. Here spectra of total (I) and polarized (P) emission  from radio surveys and WMAP data are compared with our model in the inner Galaxy, showing reasonable agreement.
We found \cite{orlando2012a, orlando2012b} that the regular field is 1.5--2 times larger than in the original models \cite{sun2008,pshirkov} based on Faraday rotation measures of extragalactic sources;
we attribute this to the existence of an anisotropic random component (also known as 'striated' or 'ordered random'), which contributes to polarized synchrotron but not to rotation measures \cite{jaffe,jansson, beck2013}.
We studied   \cite{orlando2012a, orlando2012b} the sensitivity of the synchrotron model to different formulations of B-field, CR source distribution and propagation parameters.
As an example, skymaps at 408~MHz and polarized emission at 23~GHz are shown in Fig~\ref{fig6} for a given model of B-field, halo size, and CR source distribution.
Note that local features like Loop I are not included in the model, and these account for much of the additional structure seen in the data.
The models reproduce the peak in the direction of the inner Galaxy for both total and polarized radio components.  
Figure \ref{fig5} shows some samples of latitude and longitude profiles. 

\vspace*{0.5cm}
\footnotesize{{\bf Acknowledgment:}{This work makes  use of HEALPix\footnote{http://healpix.jpl.nasa.gov/} described in \cite{healpix}.
E.O. acknowledges support via NASA Grant No. NNX13AH72G.}}


\begin{thebibliography}{}

\bibitem{diffuse2} Ackermann, M., 
Ajello, M., Atwood, W.~B., et al.\ 2012, ApJ, 750, 3  
\bibitem{jansson} Jansson, R.,\& Farrar, G. R.\ 2012, ApJ, 757, 14; ApJL, 761, 11   
\bibitem{healpix} G{\'o}rski, K.~M., 
Hivon, E., Banday, A.~J., et al.\ 2005, ApJ, 622, 759
\bibitem{beck2013} Beck, R., \& Wielebinski, R.\ 2013, Planets, Stars and Stellar Systems.~Volume 5: Galactic Structure and Stellar Populations, 641
\bibitem{jaffe} Jaffe, T.~R., Banday, A.~J., Leahy, J.~P., Leach, S., \& Strong, A.~W.\ 2011, MNRAS, 416, 1152
\bibitem{haslam81} Haslam, C.~G.~T., 
Stoffel, H., Kearsey, S., Osborne, J.~L., \& Phillipps, S.\ 1981, Nature, 289, 470 
\bibitem{moskalenko98} Moskalenko, I.~V., \& Strong, A.~W.\ 1998, ApJ, 493, 694 
\bibitem{moska2000} Moskalenko, I.~V., \& Strong, A.~W.\ 2000, ApJ, 528, 357 
\bibitem{orlando2009} Orlando, E., Strong, 
A.~W., Moskalenko, I.~V., et al.\ 2009, arXiv:0907.0553
\bibitem{orlando2012a}  Orlando, E., \& Strong, A.\ 2013, Nuclear Physics B Proceedings Supplements, 239, 64  (arXiv:1303.5488) 
\bibitem{orlando2012b} Orlando, E. \& Strong, A. W.,\ 2013, in preparation
\bibitem{porter2008} Porter, T.~A., 
Moskalenko, I.~V., Strong, A.~W., Orlando, E., 
\& Bouchet, L.\ 2008, ApJ, 682, 400 
\bibitem{pshirkov} Pshirkov, M.~S., Tinyakov, P.~G., Kronberg, P.~P., \& Newton-McGee, K.~J.\ 2011, ApJ, 738, 192 
\bibitem{strong98} Strong, A.~W., \& Moskalenko, I.~V.\ 1998, ApJ, 509, 212 
\bibitem{strong2000} Strong, A.~W., 
Moskalenko, I.~V., \& Reimer, O.\ 2000, ApJ, 537, 763 
\bibitem{strong2004} Strong, A.~W., 
Moskalenko, I.~V., \& Reimer, O.\ 2004, ApJ, 613, 962 
\bibitem{strong2007} Strong, A.~W., 
Moskalenko, I.~V., 
\& Ptuskin, V.~S.\ 2007, Annual Rev. of Nuc. and Part. Science, 57, 285 
\bibitem{strong2010} Strong, A.~W., Porter, 
T.~A., Digel, S.~W., et al.\ 2010,  ApJL, 722, L58 
\bibitem{strong2011} Strong, A.~W., Orlando, E., \& Jaffe, T.~R.\ 2011, A\&A, 534, A54 
\bibitem{sun2008} Sun, X.~H., Reich, W., Waelkens, A., \& En{\ss}lin, T.~A.\ 2008, A\&A, 477, 573 
 

\end{thebibliography}
\end{document}